\documentclass[aps,pre,twocolumn,footinbib,longbibliography,showpacs,groupedaddress,floatfix,superscriptaddress]{revtex4-2}%

\usepackage[dvipsnames]{xcolor}
\definecolor{linkcolor}{rgb}{0.3,0.3,1.0} 
\usepackage[pdftex,colorlinks=true, linkcolor= linkcolor, citecolor= linkcolor, urlcolor= linkcolor, hyperindex=true,hyperfigures=true]{hyperref} 

\usepackage{amssymb,amsmath,amsfonts}
\usepackage{graphicx}
\usepackage{float}
\usepackage{bm}

\usepackage{tikz}
\usepackage{tikz}
\usetikzlibrary{babel}
\definecolor{symCyan}{RGB}{93,204,237}
\definecolor{symRed}{RGB}{255,76,76}

\usetikzlibrary{positioning}
\definecolor{mycolorX}{RGB}{220,60,20}
\definecolor{mycolorZ}{RGB}{70,85,220}

\tikzset{
  bit/.style={circle,draw,minimum size=4pt,inner sep=0pt},
  xstate/.style={
    draw,
    rounded corners=6pt,
    fill=mycolorX!40,
    minimum width=0.5cm,
    inner sep=4pt
  },
  zstate/.style={
    draw,
    rounded corners=6pt,
    fill=mycolorZ!40,
    minimum width=0.5cm,
    inner sep=4pt
  },
  >=stealth
}

\newcommand{\drawbits}[1]{%
  \begin{tikzpicture}[scale=0.8]
    \foreach[count=\i] \c in {#1} {%
      \ifnum\pdfstrcmp{\c}{b}=0
        \node[bit,fill=black] at (0,-0.35*\i) {};
      \else
        \node[bit,fill=white] at (0,-0.35*\i) {};
      \fi
    }%
  \end{tikzpicture}%
}

\newcommand{\xstate}[4][]{%
  \node[xstate,#1] (#2) {\drawbits{#4}};
  \node[below=0.1cm of #2] {#3};
}
\newcommand{\zstate}[4][]{%
  \node[zstate,#1] (#2) {\drawbits{#4}};
  \node[below=0.1cm of #2] {#3};
}

\newcommand{\transition}[5][]{%
  \ifx\relax#1\relax
    \draw[->] (#2) -- (#3) node[midway,above] {#4}
    node[midway,below] {#5};
  \else
    \draw[->,line width=#1] (#2) -- (#3) node[midway,above] {#4}
    node[midway,below] {#5};
  \fi
}

\newcommand{\mean}[1]{\left\langle {#1} \right\rangle}

\renewcommand{\bar}[1]{\overline{#1}}
\newcommand{\bw}{{\bar{w}}}
\newcommand{\ba}{{\bar{a}}}
\newcommand{\bww}{{\bar{\bm{w}}}}
\newcommand{\bp}{{\bar{p}}}
\newcommand{\bA}{{\bar{A}}}
\newcommand{\argmax}[1]{\operatorname*{arg\,max}_{#1}}

\newcommand{\zz}{{\bm{z}}}
\newcommand{\xx}{{\bm{x}}}
\newcommand{\ww}{{\bm{w}}}
\renewcommand{\aa}{{\bm{a}}}

\newcommand{\baa}{{\bar{\bm{a}}}}

\newcommand{\DD}{\mathcal{D}}

\newcommand{\LL}{\mathcal{L}}
\newcommand{\NN}{\mathcal{N}}

\newcommand{\TT}{\mathcal{T}}
\newcommand{\XX}{\mathcal{X}}
\newcommand{\ZZ}{\mathcal{Z}}

\begin{document}

\title{Emergence of Nonequilibrium Latent Cycles in Unsupervised Generative Modeling}

\author{Marco Baiesi}
\affiliation{Department of Physics and Astronomy, University of Padova, 
Via Marzolo 8, I-35131 Padova, Italy}
\affiliation{INFN, Sezione di Padova, Via Marzolo 8, I-35131 Padova, Italy}

\author{Alberto Rosso}
\affiliation{
LPTMS, CNRS, Universit\'e  Paris-Saclay,  91405,  Orsay, France
}

\begin{abstract}
We show that nonequilibrium dynamics can play a constructive role in unsupervised machine learning by inducing the spontaneous emergence of latent-state cycles. We introduce a model in which visible and hidden variables interact through two independently parametrized transition matrices, defining a Markov chain whose steady state is intrinsically out of equilibrium. Likelihood maximization drives this system toward nonequilibrium steady states with finite entropy production, reduced self-transition probabilities, and persistent probability currents in the latent space. These cycles are not imposed by the architecture but arise from training, and models that develop them reproduce the empirical distribution of data classes more faithfully, with a clear correlation between agreement with the data and entropy production. Compared with equilibrium approaches such as restricted Boltzmann machines, our model breaks the detailed balance between the forward and backward conditional transitions and relies on a log-likelihood gradient that depends explicitly on the last two steps of the Markov chain. Hence, this exploration of the interface between nonequilibrium statistical physics and modern machine learning suggests that introducing irreversibility into latent-variable models can improve the fidelity of the generated data distribution.
\end{abstract}

\maketitle

\section{Introduction}

Unsupervised generative modeling aims to learn a probability distribution $p(\xx)$ from a dataset $\DD$ of observed configurations~\cite{mehta2019high}. A widespread strategy is to introduce latent variables $\zz \in \ZZ$ and define a joint distribution $p(\xx,\zz)$ whose marginal over $\xx$ approximates the data distribution. Classical examples include energy-based models, such as restricted Boltzmann machines (RBMs) \cite{smolensky1986information,hinton2012practical}, where one specifies an energy function $E(\xx,\zz)$ and sets $p(\xx,\zz) \sim e^{-E(\xx,\zz)}$. In such architectures, the alternation between visible and hidden layers is implemented through Gibbs sampling steps that satisfy a detailed balance. Hence, the learning dynamics is tied to an equilibrium steady state, and the latent-variable transitions are necessarily reversible. While RBMs have been successful in modeling structured data (see, e.g.~\cite{malbranke2023machine,fernandez2023designing,decelle2023unsupervised,braghetto2023interpretable}), their reliance on reversible Gibbs updates tends to limit exploration of the configuration space: once a hidden configuration $\zz$ is activated by a given $\xx$, the sampler typically returns to the same neighborhood \cite{roussel2021barriers,agoritsas2023explaining,bereux2025fast}, a behavior reminiscent of slow mixing in equilibrium Monte Carlo methods \cite{newman1999monte}.

Several other classes of latent-variable models have been developed to address limitations of equilibrium generative architectures. Helmholtz machines~\cite{hinton1995wake} and deep Boltzmann machines~\cite{salakhutdinov2009deep} introduced paired ``recognition'' and ``generative'' networks, trained, for example, by the wake-sleep phases~\cite{hinton1995wake,salakhutdinov2009deep,bornschein2014reweighted}, but their sampling dynamics follows a one-way generative path rather than an alternating interaction between visible and latent variables. Modern approaches such as variational autoencoders~\cite{kingma2014autoencoding,rezende2014stochastic} and diffusion-based generative models~\cite{sohl2015deep,song2020score,biroli2024dynamical,kamb2024analytic,sclocchi2025phase,ikeda2025speed} adopt the same structure: they specify a directed generative process and use a backward inference mechanism to propagate information in the opposite direction. In these frameworks, recognition and generation proceed along separate pathways, and the latent variables do not participate in a stochastic alternation with the visible layer.

Equilibrium dynamics and detailed balance are not universal organizing principles in natural information-processing systems. In many biological contexts, functionality relies on operating out of equilibrium: energy dissipation (but also kinetic, nondissipative features~\cite{baiesi2018life}) enables enhanced precision through kinetic proofreading~\cite{murugan2012speed,gaspard2016kinetics}, sustained directional transport by molecular motors~\cite{julicher1997modeling,parmeggiani2003phase,chou2011non}, adaptive chemotactic responses~\cite{sartori2015thermodynamics}, and sensitivities that surpass equilibrium bounds~\cite{barato2014efficiency}. This perspective 
motivates us to ask whether controlled violations of detailed balance can play a similarly constructive role in latent-variable generative models, beyond the equilibrium paradigm underlying standard architectures.

\begin{figure}[b]
\centering
\resizebox{\columnwidth}{!}{%
\begin{tikzpicture}[node distance=1.0cm] 

  \xstate{x0}{$\xx(0)$}{b,w,w,w,b,w,w,b}
  \zstate[right=of x0]{z1}{$\zz(0)$}{w,b,b}
  \xstate[right=of z1]{x2}{$\xx(1)$}{w,b,b,w,w,b,b,b}
  \zstate[right=of x2]{z3}{$\zz(1)$}{b,b,w}
  \xstate[right=of z3]{x4}{$\xx(2)$}{w,b,w,b,w,b,w,b}
  \zstate[right=of x4]{z5}{$\zz(2)$}{b,w,b}
  \xstate[right=of z5]{x6}{$\xx(3)$}{b,w,w,b,b,b,b,b}

  \transition[.8]{x0}{z1}{$\bww,\baa$}{$\bA_{\xx\zz}$}
  \transition[2]{z1}{x2}{$\ww,\aa$}{$A_{\zz\xx}$}
  \transition[.8]{x2}{z3}{$\bww,\baa$}{$\bA_{\xx\zz}$}
  \transition[2]{z3}{x4}{$\ww,\aa$}{$A_{\zz\xx}$}
  \transition[.8]{x4}{z5}{$\bww,\baa$}{$\bA_{\xx\zz}$}
  \transition[2]{z5}{x6}{$\ww,\aa$}{$A_{\zz\xx}$}

\end{tikzpicture}%
}
\caption{Markov chain dynamics alternating visible states $\xx(t)$ and hidden $\zz(t)$. This sequence has $T=3$ double steps $\{\xx\to\zz\to\}$. The Markov jump probabilities $\bA_{\xx\zz}$ of the $\xx\to\zz$ transitions are determined by parameters $\{\bww,\baa\}$ differing from the parameters $\{\ww,\aa\}$ that govern jump probabilities $A_{\zz\xx}$ of the $\zz\to\xx$ transitions.}
\label{fig:scheme}
\end{figure}

We propose a generative model whose visible/latent alternation is governed by a generic nonequilibrium Markov chain in a steady state. We introduce two independently parametrized transition matrices,
\begin{align}
A_{\zz \xx} &\equiv p(\xx|\zz)\,, \label{Azxi} \\
\bA_{\xx \zz} &\equiv p(\zz|\xx)\,, \label{Axzi}
\end{align}
which need not satisfy detailed balance. The composition of these transitions defines a discrete-time Markov process in the bipartite space $\XX \cup \ZZ$ (see Fig.~\ref{fig:scheme}).
Because no energy function is available, the stationary distribution has no closed-form expression. Nevertheless, we introduce an approximate log-likelihood gradient that provides a stable and fast-converging training rule.

A central observation of this study is that training a generative model based on generic Markov transitions never selects the reversible, equilibrium limit: the learned dynamics always violates detailed balance. Moreover, the models that reproduce the data classes most faithfully, with the lowest Kullback--Leibler (KL) divergence between generated and actual data, are those that operate far from equilibrium. To quantify this distance, we study the trajectories of the hidden states through the properties of the Markov matrix
\begin{align}
\label{Mz'z}
M_{\zz'\zz} = \sum_{\xx} A_{\zz'\xx}\,\bA_{\xx\zz}\,, 
\end{align}
and characterize its behavior in terms of entropy production, persistent probability currents, reduced diagonal occupancy, and long decorrelation times.

Interestingly, likelihood maximization is sufficient to drive the hidden-layer dynamics to self-organize into regular and irreversible cycles. This cyclic behavior is not imposed by the architecture but emerges spontaneously during training as $M$ develops strong directional flows and long-range temporal structure. Models exhibiting clearer latent-state cycles
reproduce more faithfully the empirical distribution of data classes.

These results point to a mechanism through which nonequilibrium dynamics can enhance generative modeling: cycling through hidden states reduces autocorrelation among fantasy samples and encourages a more thorough exploration of the data modes. This suggests that irreversibility may be beneficial not only for sampling, as in nonreversible Monte Carlo methods~\cite{diaconis2000analysis,bernard2009event,michel2014generalized,vucelja2016lifting}, but also for learning itself.

The remainder of the paper introduces the parametrized transition matrices defining the model (Sec.~\ref{sec:model}), derives the associated approximate log-likelihood gradient (Sec.~\ref{sec:grad}) and analyzes an example of nonequilibrium steady states that arise during training through entropy production, decorrelation times, and the structure of the latent-to-latent transition matrix (Sec.~\ref{sec:example}). We end with some conclusions (Sec.~\ref{sec:concl}).

\section{Model}
\label{sec:model}

In this paper, we focus on digital data $\xx = (x_1, \ldots, x_D)$ as a $D$ dimensional array of binary or bit variables $x_i \in \{0,1\}$. Therefore, there are $2^D$ possible data configurations in the ensemble $\XX$.
Similarly, a hidden variable $\zz = (z_1, \ldots, z_L)$ is described by its components $z_i \in \{0,1\}$ and has a size $L \ll D$.

A state $\zz$ can jump to any of the $\xx$ states via a Markov matrix $A$, defined in \eqref{Azxi}, whose element $A_{\zz\xx}$ represents the probability of jumping from $\zz$ to $\xx$. Similarly, a state $\xx$ can transition back to any of the $\zz$ states through a Markov matrix $\bA$ (Eq.~\eqref{Axzi}), whose element $\bA_{\xx\zz}$ gives the probability of jumping from $\xx$ to $\zz$. In both cases, the conditions
\begin{align}
\sum_{\xx \in \XX} A_{\zz\xx} = 1, \quad\forall \zz; \quad   \sum_{\zz \in \ZZ} \bA_{\xx\zz} = 1, \quad\forall \xx       
\end{align}
ensure the normalization of each row of the Markov matrices. We denote as $p(\xx)$ the stationary probability of visible states and as $\bp(\zz)$ the stationary probability of hidden states. Given the transition matrices $A$ and  $\bA$ they satisfy the following equations:
\begin{align}
\label{px_from_z}
    p(\xx) &= \sum_{\zz\in\ZZ} \bp(\zz) A_{\zz\xx}\,, \\
    \bp(\zz) &= \sum_{\xx\in\XX} p(\xx) \bA_{\xx\zz} \label{pz_from_x}\,.
\end{align}

Following common practice in unsupervised machine learning, we train the 
model by maximizing the log-likelihood 
\begin{align}
\label{LL}
    \LL &= \frac 1 {N_\DD} \sum_m \ln p(\xx^{(m)}) \nonumber\\
    &=\sum_\xx p_{\mathrm{data}}(\xx) \ln p(\xx) \nonumber\\
    &= \mean{\ln p(\xx)}_{\xx\in\DD}
\end{align}
 averaged in the data 
set $\DD=(\xx^{(1)},\ldots,\xx^{(N_\DD)})$.

The gradient descent to maximize it is performed on the parameters that define the matrices $A$ and $\bA$. For these Markov matrices, we chose the exponential structure typically used for RBMs:
\begin{align}
    A_{\zz\xx} & = \dfrac{e^{\sum_{\alpha,i}  x_i w_{i\alpha} z_\alpha  + \sum_i x_i a_i}}{\sum_{\xx'} e^{\sum_{\alpha,i}  x'_i w_{i\alpha} z_\alpha  + \sum_i x'_i a_i}}
    = \frac{e^{\xx\cdot(\ww\cdot\zz+\aa)}}{\sum_{\xx'}e^{\xx'\cdot(\ww\cdot\zz+\aa)}}
    \,,
    \label{Azx}\\
    \bA_{\xx\zz} & = \dfrac{e^{\sum_{\alpha,i}   z_\alpha\bw_{\alpha i} x_i  + \sum_\alpha z_\alpha \ba_\alpha}}{\sum_{\zz'} e^{\sum_{\alpha,i}   z'_\alpha\bw_{\alpha i} x_i  + \sum_\alpha z'_\alpha \ba_\alpha}}
    = \frac{e^{\zz\cdot(\bww\cdot\xx+\baa)}}{\sum_{\zz'}e^{\zz'\cdot(\bww\cdot\xx+\baa)}}
    \,.
    \label{Axz}
\end{align}
Here we denote the local biases for visible and hidden units by $\aa$ and $\baa$, 
respectively. The weights are given by two separate matrices: $\ww$ of size 
$D\times L$ for the transitions $\zz \to \xx$, and $\bww$ of size $L\times D$ 
for the transitions $\xx \to \zz$. 
Note that, unlike RBMs, 
$\ww$ and $\bww$ are not transposes of each other.

The exponential parameterization of the transition matrices allows for bitwise factorization,
\begin{align}
\label{Axz_fact}
A_{\zz\xx} = \prod_{i=1}^D
\frac{e^{x_i \left(\sum_{\alpha} w_{i\alpha} z_\alpha + a_i\right)}}{
e^{\sum_{\alpha} w_{i\alpha} z_\alpha + a_i}+1 } =\prod_{i=1}^D p(x_i|\zz).
\end{align}
Similarly, the transition probabilities from $\xx$ to $\zz$ factor as
\begin{align}
\label{Azx_fact}
\bA_{\xx\zz} = \prod_{\alpha=1}^L
\frac{e^{z_\alpha \left(\sum_i \bw_{\alpha i} x_i + \ba_\alpha\right)}}{
e^{\sum_i \bw_{\alpha i} x_i + \ba_\alpha}+1 } =\prod_{\alpha=1}^L p(z_\alpha|\xx).
\end{align}

These two factorizations enable efficient sampling of the Markov chain:
the first equation allows sampling each component $x_i$ given the hidden state $\zz$, and the second one allows sampling $z_\alpha$ given $\xx$.

\section{Gradient approximation}
\label{sec:grad}

In principle, we should evaluate the derivatives of $\ln p(\xx)$ with respect to the parameters that define the matrices $\ww$,
$\bww$, and the vectors $\aa$, $\baa$, computed on the data points $\xx$.
The stationary distribution $p(\xx)$ is the eigenvector associated with
the eigenvalue 1 of the matrix
$\sum_{\zz} \bA_{\xx'\zz} A_{\zz\xx}$.
However, expressing $p(\xx)$ explicitly in terms of parameters and
then differentiating it is too complex. We therefore introduce an approximation that restricts the dependence
on the parameters to the last two steps of the Markov chain:
\begin{align}
p(\xx) &= \sum_\zz \bp(\zz) A_{\zz\xx} = \sum_\zz \sum_{\xx'} A_{\zz\xx} \bA_{\xx'\zz} p(\xx')\,.
\end{align}
In particular, the dependence on the parameters that define $\ww$ and $\aa$
is confined to the matrix $A$ appearing in the second term of the equation,
whereas the dependence on $\bww$ and $\baa$ is limited to the matrix
$\bA$ in the third term.

To compute the derivatives of the log-likelihood $\mathcal{L}$, it is useful to first compute the partial derivatives of the two Markov matrices  with respect to their parameters,
\begin{subequations}
\label{grad-A}
\begin{align}
\label{grad-A-w} 
    \partial_{w_{j\beta }} A_{\zz\xx} &=
    A_{\zz\xx} z_\beta  \left(x_j- \mu_{x_j|\zz} \right),
    \\
    \partial_{a_{j}} A_{\zz\xx} &=
    A_{\zz\xx} \left(x_j- \mu_{x_j|\zz}\right)\,,  \label{grad-A-a} \\
 \partial_{\bw_{\beta j}} \bA_{\xx\zz} &=
    \bA_{\xx\zz} x_j  \left(z_\beta- \mu_{z_\beta|\xx}\right)\,,
\label{grad-bA-bw}    \\
    \partial_{\ba_{\beta}} \bA_{\xx\zz} &=
    \bA_{\xx\zz} \left(z_\beta- \mu_{z_\beta|\xx}\right) \,,
    \label{grad-bA-ba}  
\end{align}
\end{subequations}
where $\mu_{x_j|\zz} = \sum_{\xx'} x'_j A_{\zz\xx'}
$
is the average of component \( j \) of the vector \( \xx' \) given \( \zz \); indeed, \( A_{\zz\xx'} \) is the conditional probability of \( \xx' \) given \( \zz \). Similarly, $
\mu_{z_\beta|\xx} = \sum_{\zz'} z'_\beta \bA_{\xx\zz'}
$
is the average of component \( \beta \) of the vectors \( \zz' \) given \( \xx \). 

Thanks to the factorized exponential structure chosen for the parametrization (see \eqref{Axz_fact} and \eqref{Azx_fact}) , these two quantities have simple analytical expressions: 
\begin{align}
\label{mean_xj}
    \mu_{x_j|\zz} = \frac{1}{1+e^{- (\ww\cdot\zz+\aa)_j}}\,, \\
    \mu_{z_\beta|\xx}=\frac{1}{1+e^{- (\bww\cdot\xx+\baa)_\beta }} \,.
    \label{mean_z_beta}
\end{align}
Note that \eqref{mean_xj} depends only on $\zz$, while \eqref{mean_z_beta} depends only on $\xx$.

The gradient components of a point's log-likelihood $\ln p(\xx)$ 
with respect to $w_{j\beta}$ or $a_j$ can be computed using 
$
p(\xx) = \sum_\zz \bp(\zz) A_{\zz\xx}$,
under the approximation that $\bp(\zz)$ is independent of 
$w_{j\beta}$ or $a_j$. From Eqs.~\eqref{grad-A-w} and \eqref{grad-A-a} we have
\begin{subequations}
\label{grad-p}
\begin{align}
\label{grad-p-w} 
  \partial_{w_{j\beta}} \ln p(\xx) &=  \frac{\mean{A_{\zz\xx}\, z_\beta
    \left( x_j -\mu_{x_j|\zz}\right)}_{\bp(\zz)}}{\mean{A_{\zz\xx}}_{\bp(\zz)}} \,,
    \\
    \partial_{a_j} \ln p(\xx) &=  \frac{\mean{A_{\zz\xx}
    \left( x_j -\mu_{x_j|\zz}\right)}_{\bp(\zz)}}{\mean{A_{\zz\xx}}_{\bp(\zz)}} \,,
    \label{grad-p-a}
\end{align}
Here $\xx$ is a given point while $ \mean{.}_{\bp(\zz)}
$ denotes the average over the $\zz$ drawn from the model.
The same applies to the derivatives with respect to 
$\bw_{\beta j}$ or $\ba_j$, where one should use
$ p(\xx)=
\sum_\zz \sum_{\xx'} A_{\zz\xx} \bA_{\xx'\zz} p(\xx')$, together with Eqs.~\eqref{grad-bA-bw} and \eqref{grad-bA-ba}:
\begin{align}
\label{grad-p-bw} 
    \partial_{\bw_{\beta j}} \ln p(\xx) &=  \frac{ \mean{A_{\zz\xx}\,
    x'_j\left( z_\beta - \mu_{z_\beta|\xx'} \right) }_{p(\xx',\zz)}  }{\mean{A_{\zz\xx}}_{\bp(\zz)}}\, \\
    \partial_{\ba_{\beta}} \ln p(\xx) &= \frac{ \mean{A_{\zz\xx}\,
    \left( z_\beta - \mu_{z_\beta|\xx'} \right) }_{p(\xx',\zz)}  }{\mean{A_{\zz\xx}}_{\bp(\zz)}}.
    \label{grad-p-ba}  
\end{align}
\end{subequations}
Here, $\xx$ is a given point, while $\mean{.}_{p(\xx',\zz)}$ denotes the average over $\xx'$ and $\zz$ drawn from the model, with $p(\xx',\zz) = p(\xx')\bA_{\xx'\zz}$.

To evaluate the approximate gradient, two averages must be taken into account.
The first corresponds to the data average over the true samples $\xx$ in~\eqref{LL} applied to the approximate gradient components in~\eqref{grad-p}. This is done by randomly selecting a mini-batch $\NN \subset \DD$ of $N$ data points (with $N \ll N_\DD$). In practice, this replaces the log-likelihood of full data, $\LL$, with a stochastic estimate $\LL_\NN$ \cite{mehta2019high}.

We also adopt the so-called “centering trick”~\cite{mont12}, which replaces the data $\xx$ with their deviation from the data average, $\xx - \langle \xx \rangle_{x \in \DD}$, in the exponents of the transition matrix $\bA_{\xx\zz}$. In practice, this modifies the gradient~\eqref{grad-p-bw} into
\begin{align}
\label{grad-p-bw_ct}
&\partial_{\bw_{\beta j}} \ln p(\xx) =\nonumber\\
&\quad\frac{
\mean{
A_{\zz\xx}\,
\big(x'_j-\langle x_j \rangle_{\xx \in \DD}\big)
\left( z_\beta - \mu_{z_\beta|\xx'} \right)
}_{p(\xx',\zz)}
}{
\mean{A_{\zz\xx}}_{\bp(\zz)}
}.
\end{align}
An analogous variant that removes the averages of the hidden variables was not necessary.
For RBMs, the centering trick stabilizes learning and makes it independent of the binarization convention (${0,1}$ or ${1,0}$). We adopt this trick because we find that it reduces the chance that some hidden units of our model never activate.

The second average is taken over the model distribution, i.e., over the states $\xx'$ and the hidden variables $\zz$ obtained from steady-state sampling of the Markov chain defined by the transition matrices $A$ and $\bA$, as sketched in Fig.~\ref{fig:scheme}. In practice, using~\eqref{Axz_fact} and~\eqref{Azx_fact}, we run a Markov chain for $T$ double steps, producing the sequence
\begin{align}
\TT = [\xx'(0), \zz(0), \xx'(1), \zz(1), \ldots, \xx'(T-1), \zz(T-1)].
\end{align}
The prime notation reminds us that these $\xx'$ are “fantasy particles” generated by the model and not true data.
To ensure efficient sampling, we employ \emph{persistent contrastive divergence}.
The final fantasy state $\xx'(T)$ is reused as the initial state $\xx'(0)$ for the next Markov chain, updated with the new transition matrices.
This persistent initialization keeps the chain close to its steady state because the parameters evolve only slowly.

After computing the stochastic gradients of the log-likelihood, we update the transition matrices using a stochastic gradient-ascent step with the RMSprop method, tuned with an appropriate learning rate. A new iteration then restarts with a new mini-batch.

\subsection{Exact gradient and fixed points}
To clarify the effect of the approximation, it is useful to decompose the exact derivative of the log-likelihood into two contributions. 
Let $K_\theta \equiv \bA_\theta A_\theta$ denote the Markov operator acting on visible states (one full $\xx\!\to\!\zz\!\to\!\xx$ double step), so that their stationary distribution satisfies
\begin{equation}
p_\theta = p_\theta K_\theta .
\label{eq:fixed_point_K}
\end{equation}
     Here, for clarity, we make explicit the dependence of the operators and of the stationary distribution on the model parameters $\theta$, which is otherwise left implicit throughout the paper.
Using this fixed-point relation, the log-likelihood can be written as
\begin{align}
\LL(\theta)
&= \sum_{\xx} p_{\mathrm{data}}(\xx)\,\ln p_\theta(\xx)\nonumber\\
&= \sum_{\xx} p_{\mathrm{data}}(\xx)\,\ln\!\big[(p_\theta K_\theta)(\xx)\big].
\end{align}
Differentiating and using $p_\theta K_\theta = p_\theta$ yields
\begin{equation}
\partial_\theta \LL
=
\sum_{\xx} p_{\mathrm{data}}(\xx)
\frac{\partial_\theta (p_\theta K_\theta)(\xx)}{p_\theta(\xx)}.
\end{equation}
The derivative splits as
\begin{equation}
\partial_\theta (p_\theta K_\theta)
=
p_\theta(\partial_\theta K_\theta)
+
(\partial_\theta p_\theta) K_\theta,
\end{equation}
leading to
\begin{equation}
\partial_\theta \LL
=
\sum_{\xx} \frac{p_{\mathrm{data}}(\xx)}{p_\theta(\xx)}
\big\{
[p_\theta(\partial_\theta K_\theta)](\xx)
+
[(\partial_\theta p_\theta)K_\theta](\xx)
\big\}.
\label{eq:gradient_split_K}
\end{equation}
The approximation used in this work amounts to neglecting the second term.

In the well-specified regime where $p_\theta = p_{\mathrm{data}}$, the weights $p_{\mathrm{data}}(\xx)/p_\theta(\xx)$ reduce to unity and the two contributions in \eqref{eq:gradient_split_K} vanish separately upon summation over $\xx$. 
The first term vanishes because the derivative of the row sums of a Markov matrix is zero, implying
\begin{equation}
\sum_{\xx'} [p_\theta(\partial_\theta K_\theta)](\xx')=0.
\end{equation}
The second term vanishes because normalization of the stationary distribution implies
\begin{equation}
\sum_{\xx} \partial_\theta p_\theta(\xx)=0.
\end{equation}
Therefore, in this regime the exact and the approximate gradients share the same fixed points. Away from the optimum (or in misspecified settings where $p_\theta \neq p_{\mathrm{data}}$), the two gradients may differ, although the approximate rule provides stable and effective training in practice.

\begin{figure*}[t!]
  \centering
\includegraphics[width=0.99\textwidth]{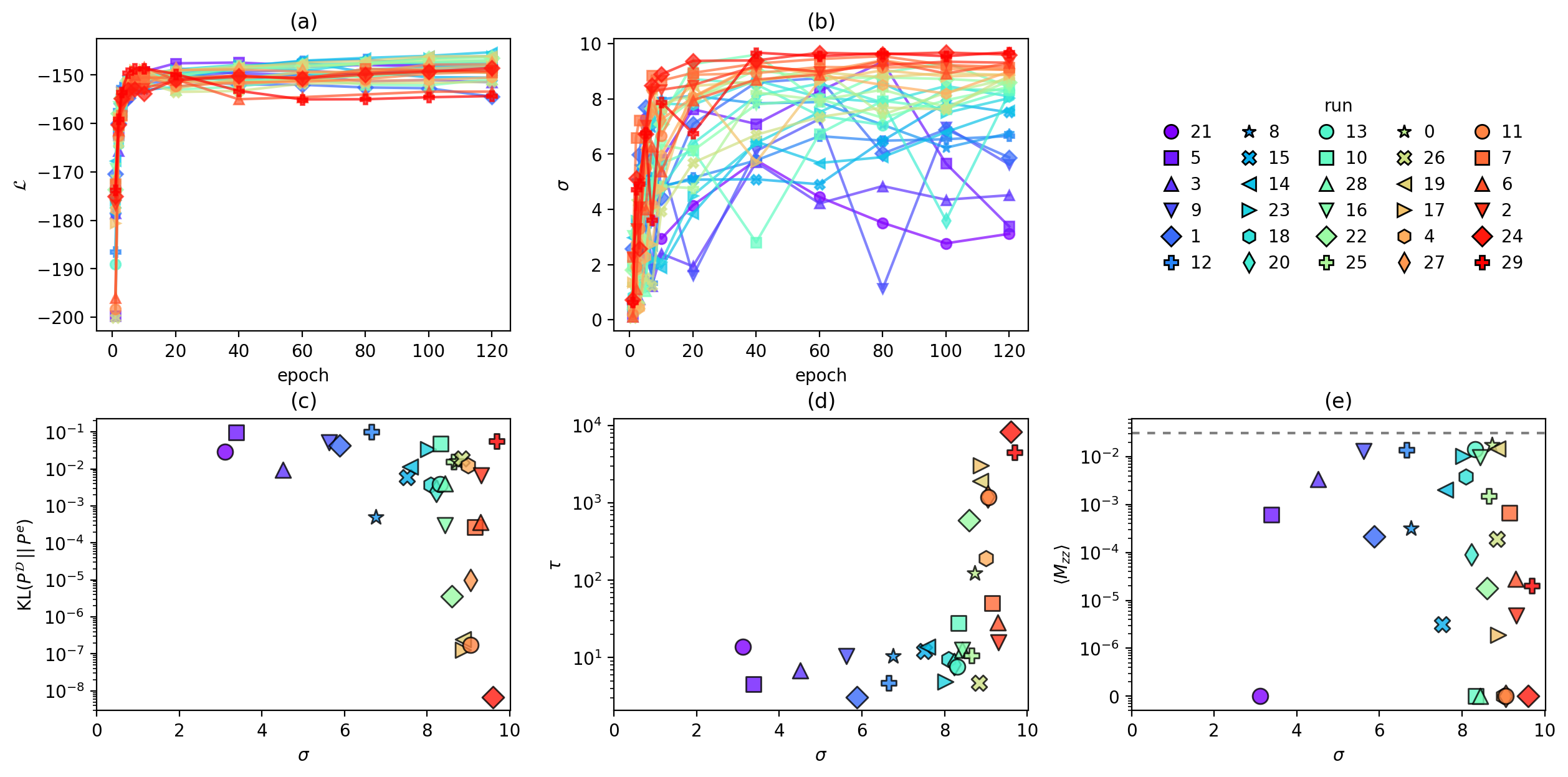}
\caption{Summary of the 30 independently trained models: (a) Log-likelihood $\LL$ and (b) entropy production rate $\sigma$ as functions of the training epoch. The color of each run encodes its final $\sigma$. The remaining panels report, at the last epoch and as functions of $\sigma$: (c) the KL divergence~\eqref{KL} between the true and generated class distributions; (d) the decorrelation time~\eqref{tau}; and (e) the mean diagonal element of the hidden transition matrix, always below the random baseline $1/2^L$ (dashed line). Low KL is achieved only at large $\sigma$.}
\label{fig:LL_S}
\end{figure*}
\begin{figure*}[t!b]
  \centering
\includegraphics[width=0.95\textwidth]{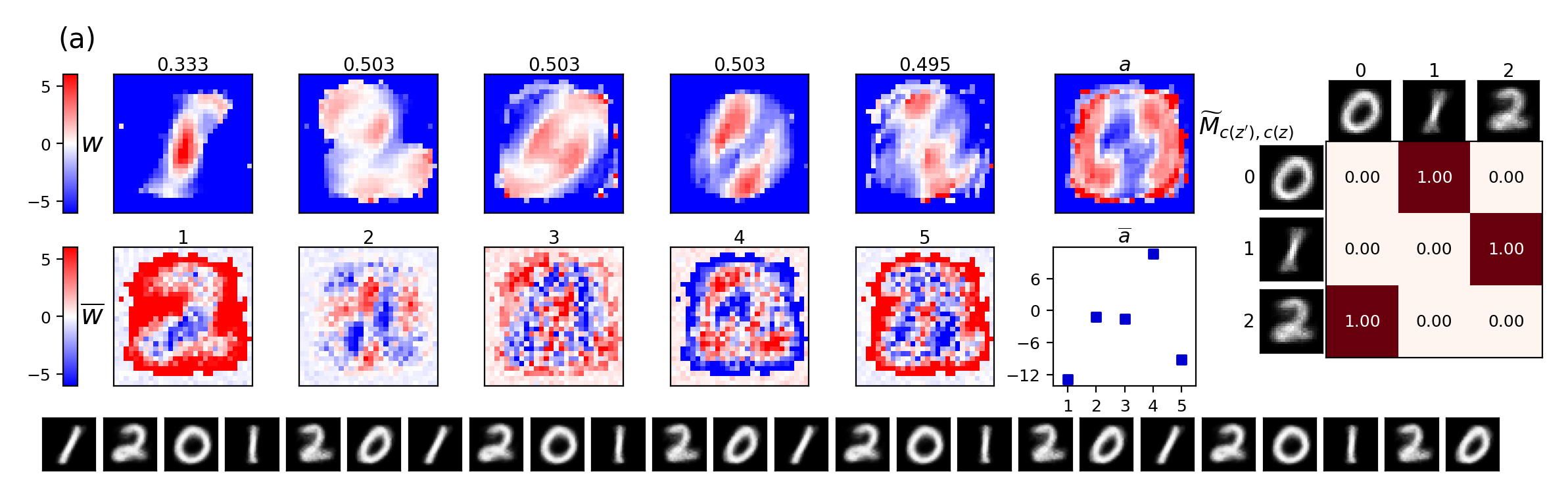}

\includegraphics[width=0.95\textwidth]{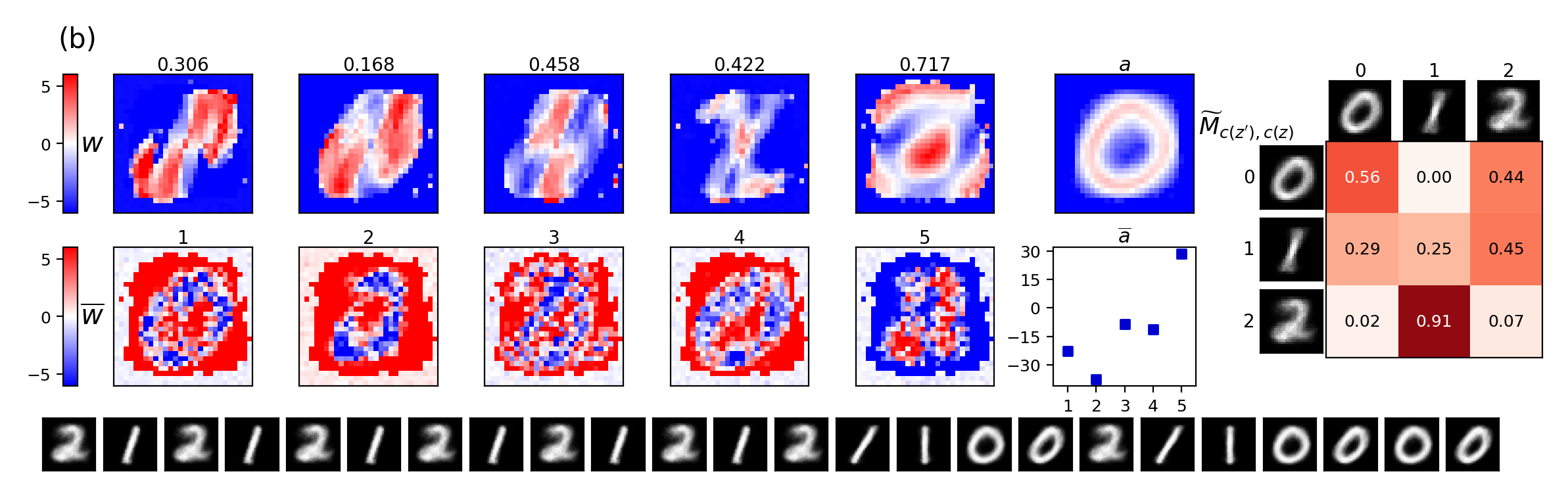}

\caption{Two representative trained models, selected for their final entropy production $\sigma$ and class-distribution divergence KL: (a) run 24, with high $\sigma$ and low KL and (b) run 3, with low $\sigma$ and high KL. Each panel shows the weights $w_{i\alpha}$ for the $L$ hidden units (first row, $D$-dimensional arrays reshaped to $28\times28$ images, with the empirical unit occupancy reported on top), the corresponding weights $\bw_{\alpha i}$ (second row), the hidden biases $\baa$ and the visible bias $\aa$ (right), and the class-projected transition matrix $\widetilde M_{c(\zz'),c(\zz)}$ between digit classes $\{0,1,2\}$ (far right; each row/column labelled by the class' representative mean image). The bottom row shows a sequence of mean generated images $\mu_{\xx|\zz(t)}$ along a sampled trajectory of hidden states. In the high-$\sigma$, low-KL run~(a) the dynamics cycles regularly $1\!\to\!2\!\to\!0\!\to\!1\ldots$, while the lower-$\sigma$ run~(b) cycles less coherently.}
\label{fig:fantasy}
\end{figure*}

\section{Example}
\label{sec:example}

\subsection{Data and parameters}

We tested the proposed unsupervised machine learning method on a selected part of the MNIST dataset of handwritten digits~\cite{lecun2002gradient}. From such visually interpretable data, we extract the first 1000 instances of the digits, 0, 1, and 2. Hence, the dataset contains $N_\DD=3000$ digits, equally balanced in three classes, and is complex yet small enough for a precise estimate of the log-likelihood.
We then digitalize each sample, assigning $x_i=1$ to pixels with a gray level above $50\%$ and $x_i=0$ otherwise. As a result, our data point $\xx^{(m)}$ is a sequence of $D=784$ bits, which can be visualized as an image by reshaping them into a matrix $28\times 28$.

We train a model with $L=5$ hidden units, using an RMSprop gradient ascent method with a learning rate $\eta=0.1$. We monitor the learning process by checking the state of the system every $30$ gradient ascent steps, a span that we term an {\em epoch}.
During the run, composed of $120$ epochs, we schedule the size of the mini-batch to increase linearly from $N=10$ to $N=50$. In each step, we run a Markov chain with $T=40$ steps. To gather statistics, we train $30$ independent models that differ only in their random initialization.

We use the general (compositional) hidden representation introduced in Sec.~\ref{sec:model}: each of the $L$ hidden units is a Bernoulli variable activated in a $\xx\to\zz$ step with probability $p(z_\alpha=1\,|\,\xx) = \mu_{z_\alpha|\xx}$ from~\eqref{mean_z_beta}, so that the hidden state $\zz$ explores all $2^L$ binary configurations. The visible variables are centered on their empirical data mean to stabilize the gradient updates.
The initialization of weights is described in the appendix, together with other technical details on stabilization of the code and computation of the log-likelihood.

\subsection{Results}

Our goal is to understand how the generative performance of the Markov-matrix architecture relates to the dynamical properties of the Markov chain it induces, and in particular to how far its stationary state lies from equilibrium.

As a first measure of generative performance, we monitor the log-likelihood $\LL$ maximized during the training of $30$ models [Fig.~\ref{fig:LL_S}(a)]. As a complementary and more stringent measure of generative quality, we also quantify how faithfully the model reproduces the class statistics of the data through the KL divergence between the true class distribution $P^\DD$ and the one generated by the model, $P^e$,
\begin{align}
    \mathrm{KL}(P^\DD \Vert P^e)
        = \sum_{k\in\{0,1,2\}}
        P^\DD(k)\ln\frac{P^\DD(k)}{P^e(k)},
    \label{KL}
\end{align}
where $P^\DD(k)=1/3$ reflects the balanced three-class dataset and $P^e(k)$ is the stationary occupancy of the hidden chain projected onto the digit classes,
\begin{align}
    P^e(k) = \sum_{\zz} \bp(\zz)\,\delta_{k,\,c(\zz)} ,
    \label{Pe}
\end{align}
with $\bp(\zz)$ the empirical stationary probability of state $\zz$ and $c(\zz)\in\{0,1,2\}$ the digit class assigned to it. The assignment compares the mean image generated by state $\zz$, i.e., the conditional average $\mu_{x_i|\zz}$ of \eqref{mean_xj}, with the empirical mean image $\mean{x_i}_{\xx\in\DD_k}$ of each class (the average over the data points of class $k$), and selects the best-overlapping one,
\begin{align}
    c(\zz) = \argmax{k\in\{0,1,2\}}\; \sum_{i=1}^D \mu_{x_i|\zz}\mean{x_i}_{\xx\in\DD_k} .
    \label{cz}
\end{align}
The same map $c(\zz)$ defines the class-projected transition matrix $\widetilde M_{c(\zz'),c(\zz)}$ shown in Fig.~\ref{fig:fantasy}. To characterize the dynamics, we then follow the trajectories of the hidden units $\zz$ and empirically estimate the transition matrix $M_{\zz'\zz}$ of Eq.~\eqref{Mz'z} by counting the transitions on long trajectories and normalizing each row. The stationary distribution $\bp(\zz)$ is likewise obtained empirically. As a diagnostic of time-reversal symmetry breaking in the nonequilibrium steady state, we compute the entropy production rate
\begin{align}
    \sigma = 
    \sum_{\zz'<\zz} 
    \left[
        \bp(\zz') M_{\zz'\zz}
        -\bp(\zz) M_{\zz\zz'}
    \right]
    \ln\frac{
        \bp(\zz') M_{\zz'\zz}
    }{
        \bp(\zz) M_{\zz\zz'}
    }.
\end{align}
Figure~\ref{fig:LL_S}(b) shows the evolution of $\sigma$ during training: the models reach comparable log-likelihoods but spread over a wide range of positive final entropy production, none approaching the reversible limit $\sigma=0$. While $\LL$ displays no systematic dependence on $\sigma$, Fig.~\ref{fig:LL_S}(c) shows that the KL divergence may become very low only as $\sigma$ grows. The benefit of operating far from equilibrium therefore emerges as a more faithful reproduction of the data classes (a lower KL).

To inspect in detail the features learned by the model, Fig.~\ref{fig:fantasy} shows two representative runs, selected by their final entropy production $\sigma$ and class KL divergence: one with large $\sigma$ and very low KL [Fig.~\ref{fig:fantasy}(a)] and one with smaller $\sigma$ and higher KL [Fig.~\ref{fig:fantasy}(b)]. Each panel displays the learned weights $\ww$, $\bww$, $\baa$ and the visible bias $\aa$, the class-projected transition matrix $\widetilde M$ between the digit classes, the per-unit average activations indicated above the $\ww$ panels, and, in the lower row, the average images generated along a sampled sequence of hidden states $(\zz(0),\zz(1),\ldots)$.

In the high-$\sigma$, low-KL run [Fig.~\ref{fig:fantasy}(a)], the hidden units capture clean, interpretable digit features and the three classes are almost equally represented. The class-projected matrix $\widetilde M$ has a dominant off-diagonal entry in every row, producing strongly irreversible trajectories that cycle regularly through the digits.
In the low-$\sigma$ run [Fig.~\ref{fig:fantasy}(b)], the dynamics displays only partial directionality and switches between short subcycles. This added stochasticity may underlie the observed increase in the KL divergence.

We confirm this picture using additional dynamical indicators. First, to quantify the regularity of the dynamics, we compute the decorrelation time
\begin{align}
    \tau = -1 / \ln |\lambda_2|,
    \label{tau}
\end{align}
where $\lambda_2$ is the second-largest eigenvalue (in modulus) of $M$.
As shown in Fig.~\ref{fig:LL_S}(d), $\tau$ correlates strongly with $\sigma$: models with large entropy production develop long correlation times, reaching up to $\tau \approx 10^4$ steps, consistent with nearly deterministic cycles.

To further assess dynamical structure, we examine the mean diagonal element of the transition matrix,
\begin{align}
    \langle M_{\zz\zz} \rangle = \frac{\mathrm{Tr}(M)}{2^L}\,.
\end{align}
Figure~\ref{fig:LL_S}(e) shows that no trained model exhibits self-transition probabilities that exceed the random baseline $1/2^L$. Even low-$\sigma$ models satisfy $\langle M_{\zz\zz} \rangle < 1/2^L$, indicating that training consistently leads to nonequilibrium steady states in which successive hidden states, and consequently the generated fantasy samples, are more decorrelated than in independent random sampling.

\section{Conclusions}
\label{sec:concl}

We introduced a generative model in which visible and latent variables interact through two independently parametrized transition kernels forming a nonequilibrium Markov chain. In comparison with classical RBM equilibrium sampling, this construction breaks the detailed balance between the forward and backward conditional transitions, which leads to qualitatively new behavior. Likelihood maximization drives the system toward nonequilibrium steady states characterized by irreversible probability currents, reduced self-transition probabilities, and the emergence of latent-state cycles. 

A quantitative comparison with standard RBMs trained on the same dataset (reported in Appendix~\ref{App:RBM}) shows that the RBMs reach lower log-likelihoods and, when used as a generative model, tend to remain trapped in a single digit for many consecutive samples. In contrast, our model naturally develops cyclic latent dynamics that promotes a regular alternation between digits.

These cycles are not imposed a priori but arise spontaneously during training. Models that develop a clearer cyclic structure exhibit higher entropy production, longer decorrelation times, and more regular alternation through latent modes. At the same time, they reproduce more faithfully the empirical distribution of classes in the dataset (with a lower KL) even though the log-likelihood itself does not single out these far-from-equilibrium solutions. This suggests that cycling improves the quality of the generative process by decorrelating successive fantasy samples and encouraging a fuller exploration of the data modes.

Our findings bring a nonequilibrium perspective to problems that are usually tackled using equilibrium RBM approaches. Several works focused on understanding and improving the slow sampling caused by the energy landscape that RBMs progressively sculpt during their learning~\cite{barra2017phase,tubi17,roussel2021barriers,agoritsas2023explaining,ventura2024unlearning,bereux2025fast}, clarifying the equilibrium and nonequilibrium regimes that arise during learning~\cite{salazar2017nonequilibrium,dece22} and the dynamical transitions that can shape training trajectories in energy-based models~\cite{bachtis2024cascade}. These approaches improve the efficiency of Monte Carlo sampling while preserving the detailed balance. In contrast, the mechanism uncovered in this work relies on explicitly breaking reversibility: the most performant models are those whose latent dynamics operates far from equilibrium and organizes into directed cycles.

Recent developments in diffusion-based generative models have highlighted similar connections between learning, sampling efficiency, and nonequilibrium thermodynamics. In particular, entropy production has been shown to constrain both the speed and accuracy of diffusion model generation~\cite{ikeda2025speed}. Our findings confirm that nonequilibrium principles can be exploited to design generative models: introducing controlled irreversibility at the latent level can improve exploration and generative performance.

In summary, this work provides a proof of principle that nonequilibrium dynamics can enhance unsupervised generative modeling, even in the minimal setting of a two-layer architecture. The learning rule derived here relies on a log-likelihood gradient that depends on the last two steps of the Markov chain, a structure absent from equilibrium energy-based models. Our results open several directions for future research, including extensions to continuous latent spaces, deeper chains, and a systematic investigation of whether nonequilibrium cycles can provide computational advantages in large-scale generative tasks.

\begin{acknowledgments}
The authors are particularly grateful to Marco Boscolo for identifying a bug in an earlier version of the code,
and would like to thank him, Manjodh Singh, Aur\'elien Decelle, Francesco La Rovere, Nicol\`o Montagner, Raffaele Sabatini, Beatriz Seoane, Peter Sollich, and Lorenzo Rosset for the many fruitful discussions. M.B.~acknowledges support for visits to LPTMS from the Centre National de la Recherche Scientifique (CNRS) and Universit\'e Paris-Saclay, and thanks LPTMS and LPTHE (Sorbonne Universit\'e) for their hospitality.
\end{acknowledgments}

\appendix

\section{Technical details}
\label{sec:app}

\subsection{Initialization of weights}
\label{ssec:A_ini}

The backward weights and biases $\bww$ and $\baa$ are set to zero
so that the hidden states $\zz$ are sampled uniformly during the early stages of learning. The forward weights $\ww$ are initialized to small independent Gaussian values,
\[
w_{i\alpha} \sim \mathcal{N}\!\left(0,\,\sigma_0^2\right),
\qquad
\sigma_0 = \frac{2}{\sqrt{D+L}},
\]
so that every hidden unit starts close to the uninformative point and the units differentiate spontaneously during learning. The visible bias $\aa$ is initialized to zero and learned together with $\ww$.

\subsection{Factorized transition probabilities}
\label{ssec:A_fact}

A numerically more stable variant replaces $A_{\zz\xx}$ by
\[
    A^*_{\zz\xx} = \prod_{i=1}^D 
    \left[\frac{e^{x_i (\sum_{\alpha} w_{i\alpha} z_\alpha + a_i)}}{e^{\sum_{\alpha} w_{i\alpha} z_\alpha + a_i}+1}\, Q\right],
\]
where $Q \approx 1.5$--$1.8$ compensates for typical sigmoid magnitudes. The ratios $A^*_{\zz\xx} / \mean{A^*_{\zz\xx}}_{\bp(\zz)}$ can then replace $A_{\zz\xx} / \mean{A_{\zz\xx}}_{\bp(\zz)}$ in gradient computations, while the log-likelihoods computed with $A^*$ include a correction term $-D \ln Q$.
In the numerical experiments, the hidden units are sampled independently from the factorized Bernoulli form~\eqref{Azx_fact}, so that the latent state explores all $2^L$ compositional configurations.

\begin{figure*}[t!b]
  \centering
\includegraphics[width=0.99\textwidth]{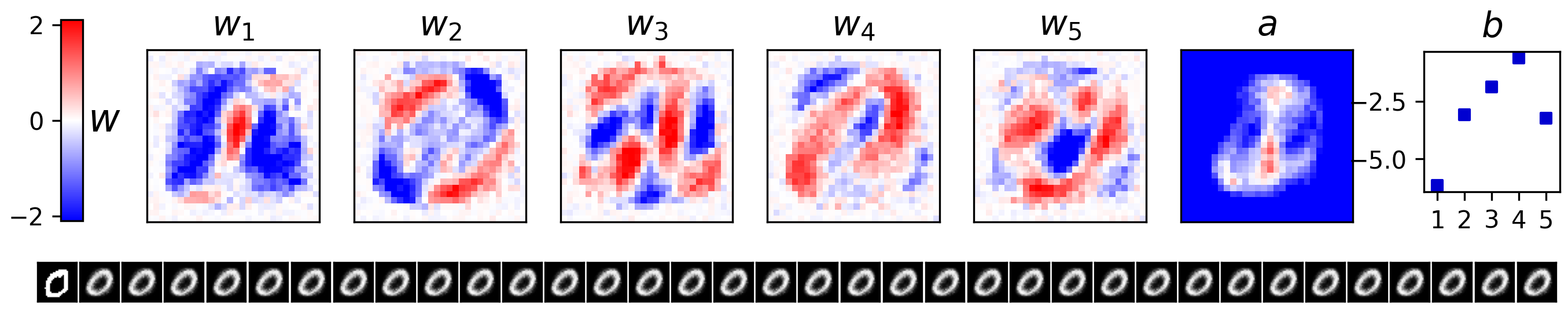}
\caption{A standard RBM trained on the same dataset with $L=5$ binary hidden units, shown for comparison with the nonequilibrium model. The first five panels display the learned weight maps $w_{i\alpha}$ of the hidden units, followed by the trained visible bias $a$ and the hidden bias $b$ (rightmost scatter). The bottom row shows a trajectory sampled with contrastive divergence, starting from a data point ``0'': the chain remains trapped in the same digit mode for the whole shown sequence, illustrating the slow mixing of RBM sampling caused by energy barriers in the learned landscape~\cite{roussel2021barriers,agoritsas2023explaining,bereux2025fast}.}
\label{fig:RBM}
\end{figure*}

\subsection{Efficient log-likelihood estimation}

A fast on-the-fly estimate of the log-likelihood for a mini-batch $\NN$ is obtained from
\[
\LL_\NN \approx \frac{1}{N} \sum_{\xx \in \NN} \ln \mean{A_{\zz\xx}}_{\zz}
\]
where the average is over sampled $\zz$ states and time steps $t$.  
A more precise estimate uses long Markov chains (e.g.\ $T=10^5$) and all $N_\DD$ data points, which requires evaluating $T\times N_\DD$ transition probabilities.  
When the latent space is small, instead one can use the empirical occupancy $\bp_e(\zz)$ obtained from the $T$ Markov samples and compute each $\ln p(\xx)$ using~\eqref{px_from_z}.  
This reduces the number of evaluations to $2^L\times N_\DD$, feasible because the latent space is small ($2^L=32$ for $L=5$).

\section{Comparison with a standard RBM} \label{App:RBM}

For comparison, we trained a standard RBM on the same dataset, with the same number $L=5$ of binary hidden units, using RMSprop together with contrastive divergence with $k=10$ Gibbs steps and the centering trick \cite{mont12}. 

The RBM defines a joint distribution over visible and hidden configurations $(\xx,\zz)$
\[
p_\theta(\xx,\zz) =
\frac{e^{-E_\theta(\xx,\zz)}}{Z_\theta},
\qquad
Z_\theta = \sum_{\xx,\zz} e^{-E_\theta(\xx,\zz)}.
\]
With binary visible and hidden units, the energy reads
\[
E_\theta(\xx,\zz)
= -\sum_i a_i x_i
   -\sum_\alpha b_\alpha z_\alpha
   -\sum_{i,\alpha} x_i w_{i\alpha} z_\alpha \,.
\]

The visible distribution is obtained by marginalizing over the hidden variables,
\[
p_\theta(\xx) = \sum_{\zz} p_\theta(\xx,\zz).
\]
The log-likelihood of the dataset is defined as
\[
\LL_\theta = \frac{1}{N_\DD}\sum_{\xx\in\DD} \ln p_\theta(\xx).
\]

Because the number of hidden units is small ($L=5$), the log-likelihood can be computed exactly by summing over hidden configurations. Taking advantage of the factorized form of the RBM energy function leads to
\[
Z_\theta =
\sum_{\zz}
e^{\sum_\alpha b_\alpha z_\alpha}
\prod_i
\left(1+e^{a_i+\sum_\alpha w_{i\alpha} z_\alpha}\right),
\]
which allows a direct evaluation of $p_\theta(\xx)$ and therefore of $\LL_\theta$.

With $L=5$ binary hidden units, the RBM reaches a best average log-likelihood of about $\LL\approx -154$, below the $\LL\approx -145$ to $-148$ attained by our nonequilibrium Markov-matrix model on the same data. A representative realization with the best log-likelihood is shown in Fig.~\ref{fig:RBM}: its bottom panel shows that, when used as a generative model, the RBM frequently becomes trapped in a single digit mode for many consecutive samples, consistent with the well-known slow mixing of RBM sampling caused by energy barriers in the learned energy landscape~\cite{roussel2021barriers,agoritsas2023explaining,bereux2025fast}. By contrast, our model cycles regularly through the digit classes (Fig.~\ref{fig:fantasy}), reproducing them more faithfully.


\begin{figure*}[t!b]
\centering
\includegraphics[width=0.99\textwidth]{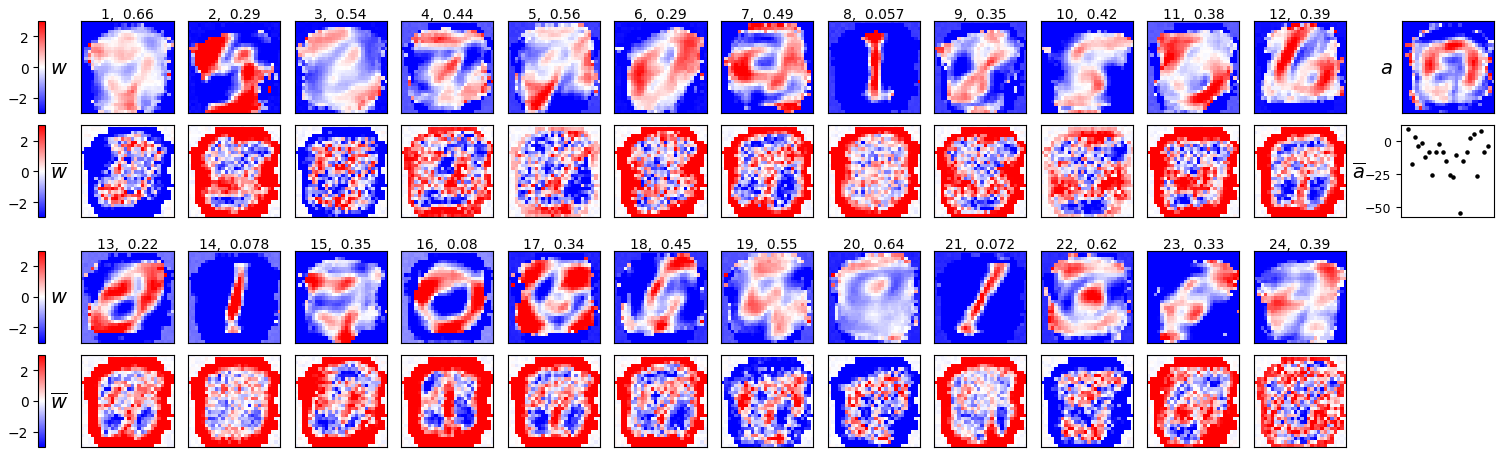}
\caption{Weights $w$ (odd rows), $\overline w$ (even rows), and visible bias $a$ and hidden bias $\overline a$ (top-right panels) learned by the model using $L=24$ hidden units and trained on MNIST.
Labels above the $w$ panels show the unit index and average activation probability in the Markov chain. 
}
\label{fig:weights24}

\vspace{3mm}
\centering
\includegraphics[width=0.94\textwidth]{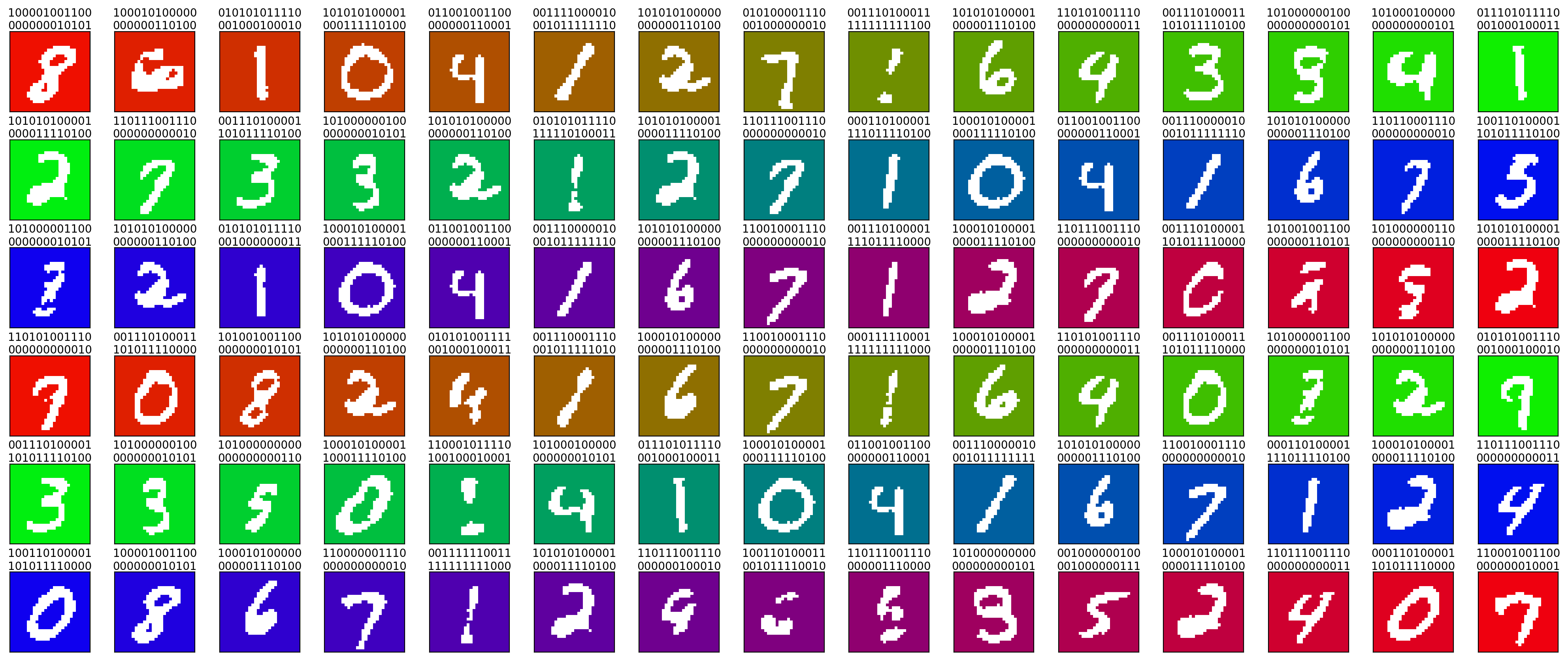}
\caption{
Markov chain generated by the model shown in Fig.~\ref{fig:weights24}. Each panel shows the visible configuration together with the hidden binary pattern that generated it, displayed above the image, which is the digitized average output of that hidden configuration. The sequence progresses row by row and illustrates how digits emerge from combinations of shared hidden features, and confirms that the model is capable of generating sequences of independent digits.
}
\label{fig:traj}
\end{figure*}

\section{Scalability test on MNIST}

To evaluate how well the model scales beyond the illustrative examples presented in the main text, we trained it on a subset of the MNIST dataset that includes all ten handwritten digit classes (using the first 500 samples from each class). The purpose of this experiment is not to reach state-of-the-art performance but rather to confirm that the learning dynamics remain stable and that the model is able to extract compositional features in a realistic, high-dimensional scenario.

The model uses $L=24$ hidden units that encode $2^{24}$ possible configurations. Its parameters are trained using RMSprop with a learning rate of $0.02$. Training involves $12000$ gradient update steps, with mini-batch sizes progressively increasing from $40$ to $240$ data samples. The Markov chains used to generate fantasy samples have length $T=50$ during the first half of training, which is then increased to $T=100$ for the second half.

Figure~\ref{fig:weights24} shows the learned parameters. Each hidden unit develops a spatial structure that resembles the features of elementary digits such as strokes or loops. 
Many hidden units are activated across multiple digits, indicating that the model learns reusable visual primitives rather than digit-specific templates.

To illustrate the generative behavior of the model, Fig.~\ref{fig:traj} shows a trajectory of the Markov chain. The binary strings displayed above each image indicate which hidden units are active in that configuration. The digit shown corresponds to the digitized average output associated with that hidden configuration, which helps visualize the corresponding feature composition.
The sequence qualitatively suggests that the dynamics operates out of equilibrium. Certain digit sequences appear very frequently (e.g, from $1\to 6\to 7$ or $1\to2\to 7$). Such directional preferences indicate the presence of probability currents in configuration space, consistent with the nonequilibrium character of the model discussed in the main text.


%

\end{document}